# Ohm's law lost and regained: observation and impact of zeros and poles

Krishna Joshi[1,2]*, Israel Kurtz[1,2], Zhou Shi[1,2,3] and Azriel Z. Genack[1,2]*

[1]Department of Physics, Queens College of the City University of New York, Flushing, New York 11367, USA

[2]Physics Program, The Graduate Center of the City University of New York, New York New York, 10016, USA

[3]OFS Labs, 19 School House Road, Somerset, New Jersey 08873, USA

The quantum conductance and its classical wave analogue, the transmittance, are given by the sum of the eigenvalues of the transmission matrix. The lowest transmission eigenvalue in diffusive media might be expected to play a negligible role in the conductance, and, in any case, to be too small to be observed. Here, we observe the lowest transmission eigenchannel in microwave waveguides, though it is orders of magnitude below the nominal noise level, and show that the transmittance is pulled down by global correlation among transmission eigenvalues and among zeros and poles of the transmission matrix. Transmission vanishes either when the energy density on the sample output vanishes at topological transmission zeros or when the longitudinal velocity vanishes precisely at the crossover to a new channel.  This lowers the conductance by an amount proportional to the modulation of the density of states. In accord with the correspondence principle, the conductance approaches Ohm's law as the number of channels increases with sample width. The exploration of the transmission matrix opens the door to a new understanding of mesoscopic transport and ultrasensitive detection techniques.

## Introduction

Ohm's law, $V = IR$, and the geometric scaling of conductance, $G \equiv 1/R = A\sigma/L$, follow from the particle diffusion model[1]. Here, $A$ is the cross-sectional area, $\sigma$ is the conductivity, and $L$ is the sample length. The quantum nature of transport[2–7] is revealed, however, in mesoscopic samples, in which the wave function is temporally coherent. These samples are shorter than the inelastic scattering length but longer than the transport mean free path, $\ell < L < L_{\text{inel}}$. Universal conductance fluctuations (UCF) are then observed with the variance of conductance nearly independent of the sample dimensions[6,8,9] as a consequence of the correlation of flux across the sample brought about by the crossing of waves of electron quasi-particles[5,8–11]. Constructive interference of waves returning to a point after traversing a loop in opposite senses in diffusive samples leads to weak localization[5,7,12–14]. Transport remains diffusive as long as the probability that the trajectory of an electron quasi-particle will loop back to a coherence length of the trajectory of one half the wavelength, $\lambda/2$, is small[5]. Once this probability exceeds unity, waves are exponentially localized within the sample and conductance falls exponentially[15–20].

Both particle and wave aspects of transport are also observed in the propagation of classical waves through random media[21,22]. The optical transmission through a wedged random sample varies inversely with sample thickness in accord with particle diffusion theory and Ohm's law, while a static speckle pattern due to the random phases of partial waves arriving at any point forms and fluctuates with frequency of the exciting wave[23]. Fluctuations and correlation of scattered flux are enhanced as the probability that a wave trajectory will cross a coherence length of the trajectory increases in the crossover to photon localization[5,9,11,18,20,24].

Landauer showed that electronic conductance is analogous to the classical wave transmittance, $T$, with $G = \frac{I}{V} = \frac{e^2}{h}T$, where $\frac{e^2}{h}$ is the quantum of conductance[3]. In the multichannel wire geometry, the transmittance is the sum of flux transmission coefficients of the $N$ channels on the incident and outgoing surfaces of the conductor, $T = \sum_{a,b}^{N} |t_{ba}|^2$, where the $t_{ba}$ are the elements of the transmission matrix (TM)[25–36]. The transmittance can be expressed as the sum of transmission eigenvalues $\tau_n$, which are the eigenvalues of the matrix $tt^\dagger$, $T = \text{Tr}(tt^\dagger) = \sum_n^N \tau_n$[10,25,26,29]. The ensemble average of the transmittance is denoted as the dimensionless conductance, $g \equiv \langle T \rangle$. Dorokhov[26] showed that in diffusive samples, in which $N/2 > g > 1$[16,17], there are $g$ "open" channels with $\tau_n > 1/e$ that carry most of the flux. The remaining transmission eigenvalues are "closed" and fall approximately exponentially with channel index, $n$[26,27,37,28,29,38,33].

Simulations in diffusive samples[39] and measurements in low-transmission graphene nanoribbons[40] show that steps are transformed into dips. However, the origins of the scaling of the mesoscopic conductance have not been explored experimentally. With classical waves, it is possible to shape the incident wave and measure the TM in static samples. We find that it is possible to measure transmission in the lowest TE many orders of magnitude below the nominal noise level of the experiment. This is a result of correlation among the TEs and between poles and zeros of the TM.

**Experimental results**

The experimental setup used to measure spectra of the TM of random dielectric waveguides is shown in Fig. 1a. The sample and experimental method are described in Methods. The square amplitude of the transmitted field for a particular configuration, frequency, and orientation of the source and receiver antennas at all points on the measurement grid on the output surface is shown in Fig. 1a. Since the spacing between points is less than $\lambda/2$, the continuous field can be recovered via the 2D sampling theorem[41]. The continuous intensity pattern on the sample output for the same incident field (Fig. 1a) is shown in Fig. 1b. The continuous field on the sample surfaces is fit by a superposition of waveguide modes. This enables a transformation of the matrix of field transmission coefficients between points on the input and output to the TM with a basis of the propagating modes of the empty waveguide.

Averaging over the transmittance measured for random configurations in three ensembles with identical composition but different lengths give the modulated spectra of $g$ shown in Fig.

1c. Since $N/2 > g > 1$, the wave is diffusive for all sample lengths shown. The vertical dashed line at 14.8317 GHz represents the crossover from $N = 61, 62$ to $N = 63, 64$ channels. The new channels that enter are the degenerate transverse electric and transverse magnetic modes of the empty waveguide (Supplementary Note 1). Spectra of the transmittance for three random sample realizations with $L = 23$ cm and the variance of the transmittance over the ensemble measured are shown in Extended Data Figs. 1a, b. Simulations of diffusive waves in Extended Data Fig. 1c show UCF with $\text{var}(T) \sim 2/15$ [8,11].

**Fig. 1 | Measurements of conductance.**

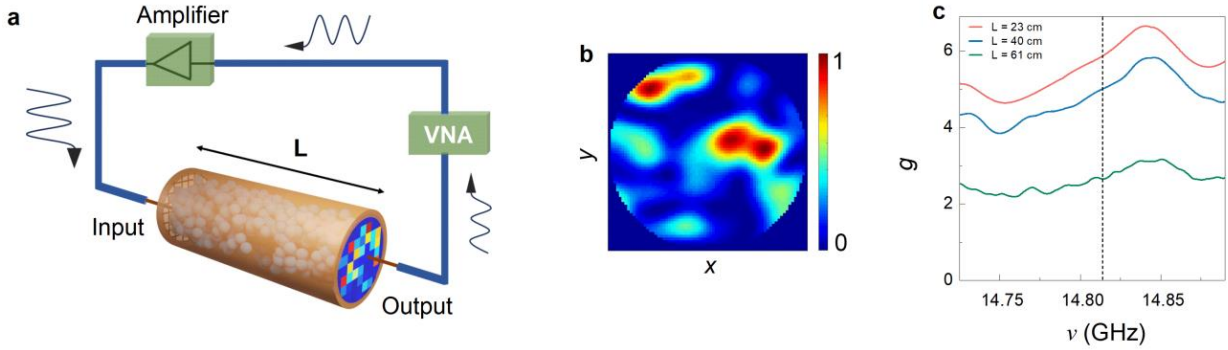

**a,** Microwave experiments are performed with use of a vector network analyser (VNA). The square of the field measured on all points on a grid on the sample output for a single position and polarization of the source antenna is shown. **b,** The 2D sampling theorem[41] is used to find the continuous field, whose square amplitude is shown. **c,** Measurements of spectra of the average transmittance, $\langle T \rangle \equiv g$, over 23 configurations for $L = 23$ cm, and 6 configurations for $L = 40$ and 61 cm.

The source of the dips in conductance is revealed in the spectra of the individual transmission eigenvalues and their factors: the longitudinal energy density at the sample output, $u_n(L)$, and the longitudinal eigenchannel velocities (EVs), $v_n(L)$[42],

$$\tau_n = u_n v_n. \qquad (1)$$

The reference to the sample output at $L$ is suppressed in the variables on the right-hand side of equation (1) to simplify the notation. Here, $u_n$ is proportional to the sum of the squares of the magnitude of electric field of the transmission eigenchannels over the output face of the sample, while $v_n$ is the average of the group velocities of waveguide modes in the transmission eigenchannel (TE) weighted by their square amplitude. The velocities of TEs are not randomized[42] and the ensemble averages of the EVs decrease as $\tau_n$ decreases, as seen in Fig. 2h.

Spectra of each of the variables in equation (1) are plotted for a single configuration of length 23 cm in Figs. 2a-d and for the average over 23 configurations in Figs. 2e-h. Transmission spectra highlighting open and closed channels, respectively, are displayed in linear and semi-logarithmic plots in Figs. 2a, b. A series of precipitous dips in $\tau_{N=62}$ before the crossover from $N = 61, 62$ to $N = 63, 64$ is shown, together with a sharp rise from ultra-low transmission precisely at the crossover for $N = 63, 64$ (Fig. 2b). The smallest value of transmission in the spectrum of $\tau_N$ is a factor of $10^9$ below the transmission in $\tau_1$ of nearly unity, while the signal to noise ratio determined from the repeatability of measurements on the time scale needed to complete the measurement of the TM for a single configuration of 40 hours is 1300 (Supplementary Note 2 and Fig. S3). Measurements of dips in transmission well below the experimental noise level reflect correlation among the transmission eigenvalues. The ability to measure the depths of the dips in $\tau_N$ is limited by the frequency step in the measurement of 300 kHz and by residual absorption. The sharpness of the dips in spectra of the lowest transmission eigenvalue suggests that they might be associated with recently predicted transmission zeros (TZs)[43,44], which have not been observed in experiments previously.

**Transmission zeros**

Transmission is predicted to vanish whenever zeros of the determinant of the TM appear on the real axis of the complex frequency plane. The determinant of the TM can be expressed as $\det(t) \sim \frac{\prod_{i=1}(\omega - \eta_i)}{\prod_{m=1}(\omega - \lambda_m)}$, where the $\eta_i = \omega_i \pm i\zeta_i$ are the zeros of $\det(t)$ and $\lambda_m = \omega_m - i\gamma_m$ are the poles[44]. The poles indicate the resonances or quasi-normal modes of the medium at frequency $\omega_m$ and half linewidth $\gamma_m$. Like the poles, the TZs are singularities in the map of the phase of the determinant of the TM, $\arg(\det(t))$, in the complex frequency plane. However, unlike the poles, which may appear anywhere in the lower half of the complex plane for unitary media, the TZs are constrained to appear either on the real axis or as complex conjugate pairs in the complex frequency plane[43]. In media in which the field decay rate due to absorption or gain, $\pm\gamma$, is uniform, the entire phase map is shifted by $\pm\gamma$, so that TZs can be shifted to the real axis by adding absorption or gain.

The spectrum of $\log \tau_{N=62}$ before the crossover from $N = 61, 62$ to $N = 63, 64$ channels (Fig. 2b) is mimicked by the spectrum of $\log u_N$ (Fig. 2c). This is consistent with equation (1) since the spectrum of $v_{N=62}$, shown in Fig. 2d, does not exhibit singular behaviour before the crossover. A quantitative check of the agreement of measurements with equation (1) is made by comparing $\tau_{n=N=62}$ found from direct analysis of the measurement of the TM in Fig. 2b with the product of $u_{62}$ and $v_{62}$ in Figs. 2c, d. These are found, respectively, from the square magnitude of the transmitted field and from the weighted average of waveguide velocities for the $62^{\text{nd}}$ TE. The plot of $\tau_N$ determined from the right-hand side of equation 1, shown as the gold circles in Fig. 2b, overlaps the curve determined from the direct analysis of the TM, confirming equation (1) and the accuracy of the determination of the parameters in the equation.

In addition to the sharp drops in $\tau_{62}$ before the crossover, the transmission eigenvalues of the new TEs that enter at the crossover vanish precisely at the crossover to $N = 63, 64$ (Figs. 2b, f). $\tau_{63}$ and $\tau_{64}$ rise sharply after the crossover as $v_{63}$ and $v_{64}$ increase from zero, while $u_{63}$ and $u_{64}$ enter with values above those for $u_{61}$ and $u_{62}$ before the crossover, as seen in Fig. 2c. Even though the contributions of closed channels to the transmittance are minimal, the appearance of zeros in transmission before and at a crossover is correlated with dips in the open channels, and so with dips in the conductance.

**Fig. 2 | Measurements of transmission eigenvalues and their factors near a crossover to a new channel.**

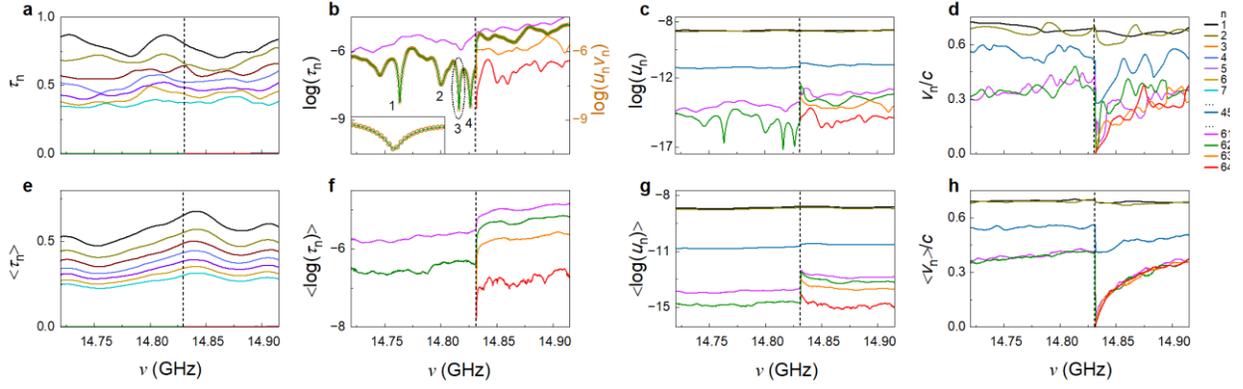

The dashed vertical lines in **(a-h)** are at the crossover from $N = 61, 62$ to $N = 63, 64$ at $v_0 = 14.8317$ GHz. **a,b,** Linear **(a)** and semi-logarithmic **(b)** plots of spectra of $\tau_n$ for a single random configuration, showing, respectively, open and closed transmission eigenvalues. The sharp dips in the spectrum in green of $\tau_{N=62}$ before the crossover are associated with TZs. An analysis below and in the Supplementary Note 4 of the transmission time in Fig. 3f shows that dips 1, 3, and 4 are due to single zeros, while dip 2 is due to a complex conjugate pair of zeros. The gold circles are the logarithm of the product of measurements of $u_{62}$ and $v_{62}$, as shown in **(c)** and **(d)**. **e-h**, The configurational average values of $\tau_n$, $\log \tau_n$, $\log u_n$, and $v_n$. Only the dips in $\log \tau_{63}$ and $\log \tau_{64}$ and $v_{63}$ and $v_{64}$ at the crossover survive averaging over configurations.

Further differences in the behaviour of zeros in transmission before and at the crossover emerge in a comparison of spectra of the properties of TEs in a single configuration (Figs. 2a-d), with the corresponding spectra of the ensemble averages (Figs. 2e-h). The sharp dips in $\tau_n$ before the crossover in single configurations disappear in the ensemble average, showing that the frequencies at which dips occur are random. However, the vanishing of $v_{63}$ and $v_{64}$ and the associated sharp rises in $\tau_{63}$ and $\tau_{64}$ precisely at the crossover survive averaging. We note that in some configurations sharp dips appear in $\tau_{64}$ after the crossover (Fig. S1). This indicates that the density of TZs is relatively high just before the crossover but does not entirely vanish after the crossover. The density of TZs becomes more uniform as the sample length increases, as seen in measurements (Fig. S1) and simulations (Fig. S2).

The identification of the dips in $\tau_N$ and $u_N$ (Figs. 2b, c) with TZs is supported by the similarity of these spectra to simulated spectra (Figs. 3a, b) and their correspondence to TZs in the plot of $\arg(\det(t))$ in the complex frequency plane (Fig. 3d). The recursive Green's function simulations used here are discussed in Methods[42,45] and a model of the sample is shown in Extended Data Fig. 2. Each of the dips in $\tau_7$ and $u_7$ in Figs. 3a, b, respectively, before the crossover from $N = 7$ to $N = 8$, corresponds to a zero on the real axis of the phase map of $\det(t)$ in the complex frequency plane shown in Fig. 3d. The zeros on the real axis are at the centres of the white circles and only occur before the crossover. A few of the poles are indicated with pink squares. The poles lie in the lower half of the complex plane and occur over the entire frequency range. EVs in the simulations shown in Fig. 3c behave similarly to the measurements in Fig. 2d: EVs only vanish at the point at which a new channel emerges. The vanishing of the EV at the crossing to a new channel produces a null in transmission which is not associated with a TZ. The crossover to the 8$^{\text{th}}$ channel in the simulations is set as the frequency of the 8$^{\text{th}}$ step in the transmittance of the empty waveguide. The crossover is not a multiple of $\lambda/2$ because of the discretization of the sample in the simulations. As the resolution in the simulation increases, however, the $N^{\text{th}}$ crossover approaches $N/(\lambda/2)$ (Methods and Extended Data Fig. 3).

In addition to the singularity in $\tau_N$ at a TZ, the transmission time $\tau_T$ in a unitary medium is singular when infinitesimal absorption or gain is added. This can be used to test whether dips in $\tau_N$ correspond to TZs. $\tau_T$ is the sum of the transmission time of all the TEs, $\tau_T = \sum_n^N t_n$ [44]. $\tau_T$ is the sum of Lorentzian functions associated with poles, $\tau_p$, and zeros, $\tau_z$, respectively, at angular frequencies $\omega_m - i\gamma_m$ and $Z_i + i\zeta_i$,

$$\tau_T = \frac{d(\arg(\det(t)))}{d\omega} = \tau_p + \tau_z = \sum_m \frac{\gamma_m}{(\omega - \omega_m)^2 + \gamma_m^2} + \sum_i \frac{\zeta_i}{(\omega - Z_i)^2 + \zeta_i^2}. \quad (2)$$

When there is no dissipation, the transmission time is proportional to the DOS and is the sum over Lorentzian lines associated only with the poles, $\tau_T = \tau_p = \sum_m \frac{\gamma_m}{(\omega - \omega_m)^2 + \gamma_m^2} = \pi\rho$ [44,46]. This requires that the contribution of the zeros to the transmission time vanishes in unitary media. Equation (2) then requires that the zeros either lie on the real axis, $\zeta_i = 0$, or be complex conjugate pairs. This creates dips (peaks) in $\tau_T$ even at the smallest level of absorption (gain) as TZs on the real axis move into the lower (upper) half of the complex frequency plane.

**Fig. 3 | Comparison of simulations of transmission eigenvalues and transmission zeros with measurements of transmission time.**

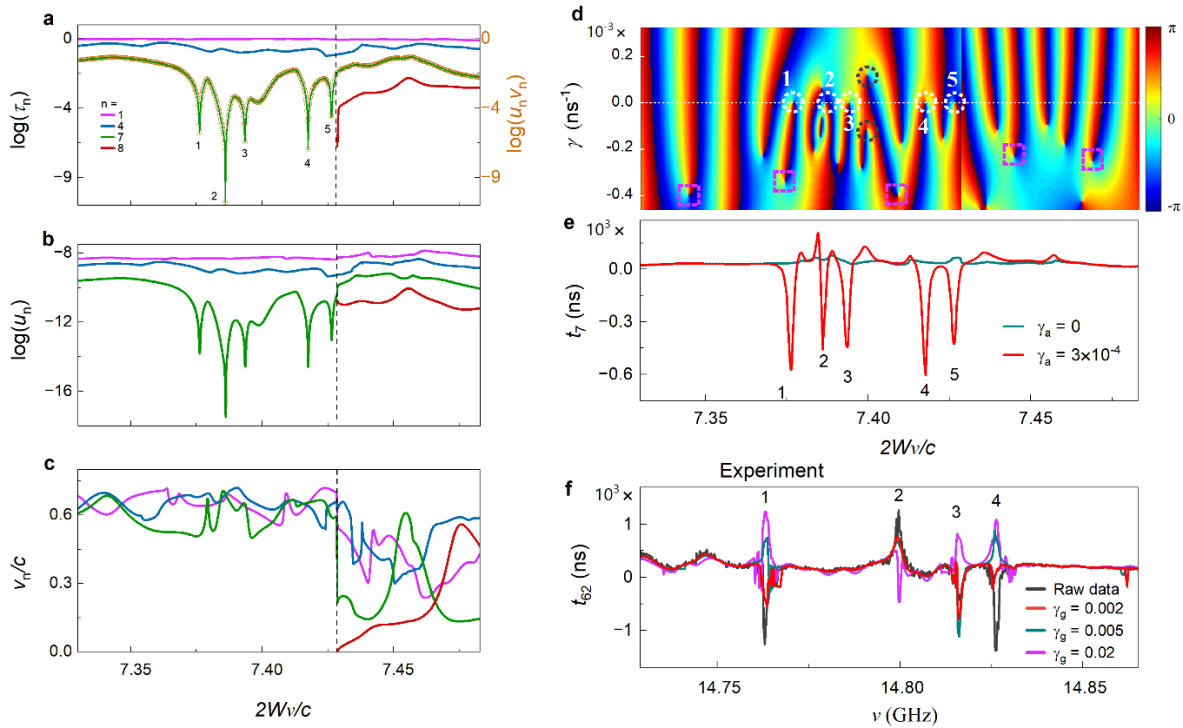

Simulations of propagation in a medium of width $26a$ and length $1000a$ near the crossover from $N = 7$ to $N = 8$ channels. Here, $a$ is the length of the sides of the square elements of the sample. **a,** Five sharp dips appear in spectra of $\log \tau_7$ before the crossover to the 8$^{th}$ propagating channel. The gold circles are the logarithm of the product $u_7 v_7$, with the factors shown in (**b, c**). The overlap of their product with $\log \tau_7$ is predicted in equation 1. **b,** Spectra of the logarithms of the linear energy density at the sample output, $u_n$, and the transmission eigenvalues, $\tau_n$, are similar except at the crossover. **c,** The EV of the new channel vanishes at the crossover and pulls down $\tau_8$ and the other EVs. **d,** Map of $\arg(\det(t))$ in the complex frequency plane. The horizontal axis is plotted in the dimensionless units $2W\nu/c = 2W/\lambda$, while the vertical axis represents the amplification rate of the field. Circles of white dots surround the TZs responsible for the five dips labelled in (**a**), while black dots circle the TZs of a conjugate pair of zeros. A few of the poles are indicated with pink squares. **e,**

Transmission time $t_n$ in a medium without absorption, $\gamma_a = 0$ (blue curve), and with absorption $\gamma_a = 3 \times 10^{-4}$ ns$^{-1}$ (red curve). **f,** The eigenchannel transmission time for $N = 62$ based on the raw data for the configuration shown in Fig. 2a-d and for three levels of added gain $\gamma_g$. The variation of $t_{62}$ with loss and gain near the frequencies of the dips in transmission in Fig. 2b is consistent with the impact of TZs on the transmission time, as given in Equation (2).

The contributions of TZs to $\tau_T$ at low levels of absorption are fully represented in $t_N$[44]. The poles that contribute to $t_N$ are far off resonance[47] and thus contribute only a flat background in the spectrum of $t_N$. Consequently, the spectral features that appear in $t_N$ when low levels of gain or loss are introduced are due to zeros of $\det(t)$ near the real axis in the complex frequency plane. The dips in $\tau_N$ in unitary media in Fig. 3a correspond to dips in $t_N$ in Fig. 3e when small absorption is added. Each of the dips in $t_N$ in Fig. 3e in samples with small absorption are nearly Lorentzian functions with depth $1/\gamma_i$ and half-width $\gamma_i$, as set forth in Equation (2) and shown in Supplementary Note 4 and Fig. S5. The different depths of the dips are due to the different average field absorption rates for different quasi-normal modes in media with nonuniform absorption. Sharp features in the experimental spectrum of $t_{62}$ in Fig. 3f appear at the same frequencies as the dips in transmission for $\tau_{62}$ for the configuration shown in Fig. 2b. Adding gain numerically to the field spectra from which the TM is constructed modifies the spectrum of $t_{62}$ in a manner consistent with a TZ near the real axis moving vertically in the complex plane.

**Global correlation**

The association of dips in the conductance near channel crossovers with reduced transmission in the lowest TE is further supported in a comparison of spectra of $g$ with spectra of the probability density of TZs, $\rho_0$, and of the average of the velocities of the lowest TE, $v_N$, in the crossovers to the 10$^{th}$ and 11$^{th}$ channels shown in Fig. 4a. $\rho_0$ is peaked right before the crossovers, while dips in $v_N$ occur after the crossover. The relatively sharp drop in $g$ before the crossover and its slower recovery after the crossover are shown to correspond to the rise in $\rho_0$ before the crossover and the slower recovery of $v_N$ from zero after the crossover (second panel in Fig. 4a).

Beyond the qualitative associations of the dips in $g$ with characteristic features of the lowest TE, there is a striking quantitative relation between dips in $g$ (top and bottom panels) and peaks in the DOS, $\rho$ (third and bottom panels in Fig. 4a). The dips in $g$ are replicas of the peaks in $\rho$ (bottom panel). Expressing $g$ and $\rho$ as sums of a slowly varying term and a remainder term with fluctuations on the scale of the change in width or frequency at which a new channel is introduced, $g = \bar{g} + \Delta g$, and $\rho = \rho_{\text{lin}} + \Delta\rho$, respectively, the remainder terms proportional to one another:

$$\Delta g = -\alpha \Delta\rho \qquad (3a)$$

$$g = \bar{g} - \alpha \Delta\rho \qquad (3b)$$

The slope of $\bar{g}$ is the average of the slopes of lines passing through the local minima and maxima of the respective functions. An excellent fit of the right-hand side of equation 3b to $g$ in terms of the parameter $\alpha$ is shown in the upper panel of Fig. 4a. For a given sample length and scattering strength, the proportionality constant, $\alpha$, is independent of sample width even as transport crosses over from being diffusive to being localized with $g$ falling below unity (Extended Data Fig. 4). The measured modulation is shown in Extended Data Fig. 5.

**Fig. 4 | Impact of transmission zeros, zeros of eigenchannel velocities, and the DOS on the dimensionless conductance.**

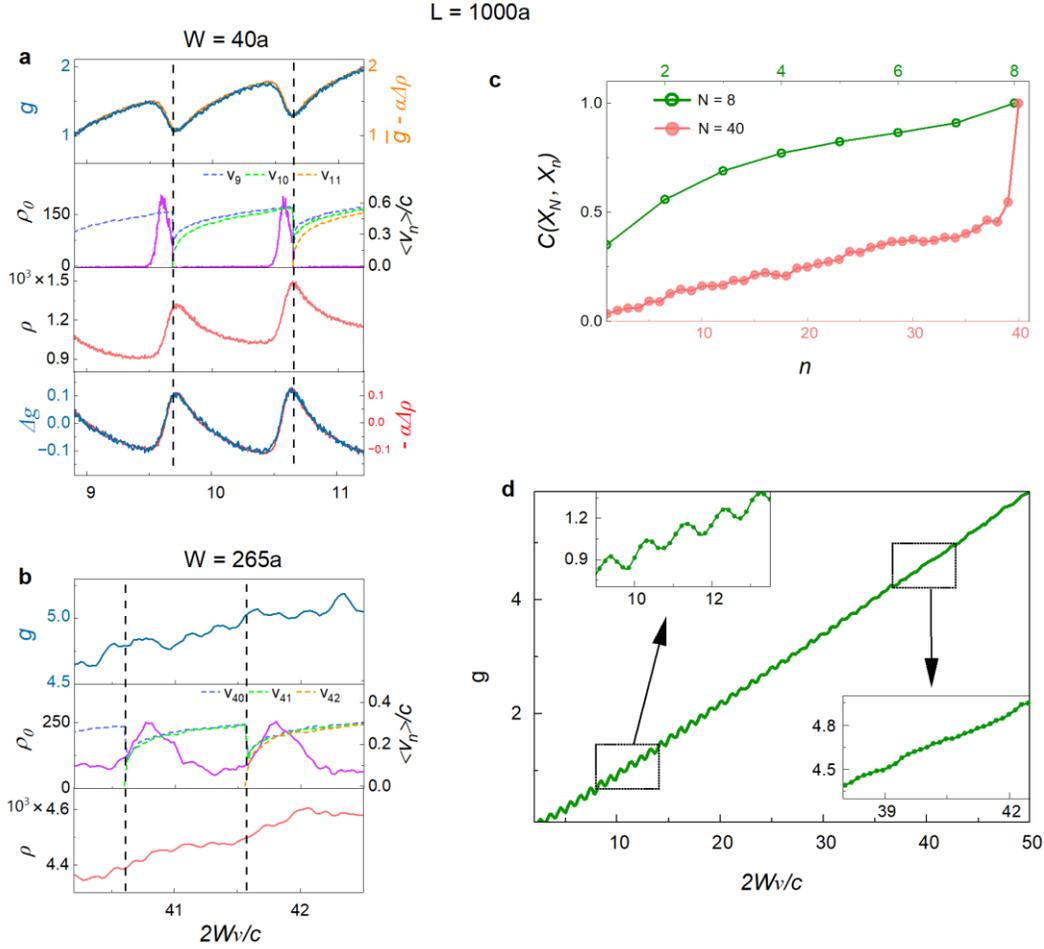

**a,** $W = 40a$; **b,** $W = 265a$, with $L = 1000a$. Simulations of spectra of $g$, the density of TZs, $\rho_0$, the EVs of the lowest transmission eigenchannel, $v_N$, the DOS, $\rho$, and the modulation of $g$ and $\rho$ vs. $2Wv/c$ in the vicinity of two channel crossovers for two ranges of sample width. The DOS is found from $\rho = \tau_T/\pi$. Spectral regions with a high density of TZs or low values of $v_N$ are seen to correspond to dips in $g$ and peaks in $\rho$. $g$ in **(a)** is fit by the difference between a slowly varying term and a term proportional to the modulated part of $\rho$, as explained below. The modulation of the DOS is diminished in wider samples, as seen in **(b)**. The peaks in $\rho_0$ shift towards lower frequencies with respect to the crossover as the sample width increases. This shift is seen in the measured conductance in Fig. 1c. **c,** Degree of correlation between the lowest transmission eigenvalues and open channels. Eigenvalues $\tau_n$ are parametrized in terms of variables $x_n$, $\tau_n \equiv \cosh^{-2} x_n$ [26]. The spacing between the $x_n$ are approximately constant for the first $N/2$ transmission eigenvalues [38]. **d,** $g$ as a function of sample width at a frequency of 14.7 GHz. The rectangular insets show zoomed-in views of the curve with strong and weak modulation of $g$ for narrow and wide samples, respectively.

The impact of TZs and low values of $v_N$ upon the conductance in samples of the same length but more than six times the width as in Fig. 4a are seen in Fig. 4b to be greatly attenuated. This reflects the weakened correlation between the open and closed transmission eigenvalues (Fig. 4c and Extended Data Fig. 6). In order to present a broad range of transmission eigenvalues, the eigenchannels $\tau_n$ are parametrized in terms of variables $x_n$, with $\tau_n \equiv \cosh^{-2} x_n$, which are approximately equally spaced for $n < N/2$ [26,38,29]. In a sample with $N \sim 8$, correlation between $x_N$ and $x_1$ is substantial, whereas it is small for $N \sim 40$. The modulation of $g$ is also reduced because TZs no longer arise exclusively just before the crossover but spread over the entire frequency range as the residual peak in $\rho_0$ broadens. As a result of the decrease in correlation among transmission eigenvalues and the widening of the range of TZs, the modulation of $g$ is washed out in the limit of large $N$. The conductance then approaches a linear, Ohmic variation with sample width (Fig. 4d).

**Discussion**

Instead of the linear scaling of the conductance with sample width predicted by Ohm's law, the measured transmittance in diffusive random waveguides is punctuated by dips. These dips are associated with vanishing transmission in the lowest TE, which are modulated with increasing frequency or sample width as new channels are introduced. The frequencies of spectrally isolated zeros of the TM coincide with dips in the transmittance (Extended Data Fig. 7), while more generally, exceptionally low transmission in the lowest TE of small samples pulls down all of the transmission eigenvalues, and so the transmittance. The correlation among low- and high-transmission eigenchannels falls as $N$ increases. In addition, the probability density of TZs broadens, shifts spectrally, and develops a plateau as the sample width increases. As a result, the scaling of the transmittance approaches Ohm's law, which follows from the particle diffusion model, and yields a linear increase in conductance with $N$ in accord with the correspondence principle.

Ultra-low transmission is found in an analysis of the measured TM at a level of up to nine orders of magnitude below the experimental noise level. Since the field in the medium, including at frequencies at which it may vanish, is determined by the constellation of poles in the map of the phase of the determinant of the TM in the complex plane, the poles and zeros are correlated. This correlation produces a dip in the conductance proportional to the rise of the DOS relative to a nearly linear increase in both these functions as the sample width increases, as given in Equation (3).

In addition to the coincidence of the start of the rise in the density of TZs, $\rho_0$, with the beginning of the dip in conductance, departures of $\rho$ and $g$ from a linear increase with frequency are proportional, $\Delta g = -\alpha \Delta \rho$. The proportionality constant, $\alpha$, is independent of sample width and falls with increasing sample length. This stands in contrast to the equivalence of $g$ and the Thouless conductance, $g_{\text{Th}}$, which is proportional to the Thouless number, $\delta$, which is the average number of modes within the modal linewidth, $\delta \omega$, $g = g_{\text{Th}} \sim \delta = \rho(\omega)\delta\omega$[16], where $\delta\omega$ is independent of width. Since $\alpha$ is independent of the sample width, $W$, for a given $L$, while $\Delta \rho$ falls with $W$, $\Delta g$ falls with $W$. This is in contrast to $g_{\text{Th}}$, which is proportional to $\rho$. Thus, the two terms in $\bar{g} - \alpha\Delta\rho$ scale differently, with $\Delta g$ tending to vanish as the sample size grows. This suggests that $\bar{g}$ may be associated with $g_{\text{Th}}$. In a full study of the scaling of conductance starting with small systems, the correction to $g$ due to the interaction of zeros and poles is crucial.

In the absence of dissipation, a conjugate pair of TZs may approach and merge on the real axis at a zero point (ZP) to produce two single TZs that subsequently move on the real axis when the sample is modified[43]. The distance between singularities in the complex plane has a square-root singularity versus modifications of the sample as a ZP is approached[43]. Observations of TZs as a sample is perturbed may therefore provide a means of ultrasensitive detection of changes in a sample.

The vanishing of the longitudinal component of velocity of the lowest-transmission eigenchannel, $v_N$, at a crossover is key to understanding the transition to a new channel. At the crossover, $v_N$ vanishes, as would be expected for the velocity of waves in an empty waveguide, as the propagation constant of an evanescent wave gains a small real component. A quantitative study of the vanishing of the EV and a peak in the DOS at channel crossovers in a random medium may shed light on the Wigner cusp in nuclear scattering at energies at which new scattering channels open[48,49] and grating anomalies[50].

**Data Availability**

The datasets generated during and/or analysed during the current study are available from figshare at XXX.

**Code Availability**

The simulation codes used in the current study are available from the corresponding author on reasonable request.

**Materials and Methods**

**Experimental setup, sample, and measurements**

The experimental setup used to measure the spectra of the TM of random dielectric waveguides is shown in Fig. 1a. The samples are collections of randomly positioned alumina spheres of diameter 0.95 cm and refractive index 3.14 embedded in Styrofoam shells to produce an alumina volume fraction of 0.07 within a cylindrical copper tube of diameter 7.3 cm. Spectra of the in- and out-of-phase components of the transmitted phase relative to the incident field are obtained for samples of lengths, $L = 23, 40,$ and 61 cm with use of a vector network analyzer (VNA) over the frequency range of 14.70-14.94 GHz. The output of the VNA is amplified to a power of nearly 1 W. Field spectra are measured for each of two perpendicular orientations of the source and receiver antennas for each pair of locations on a 9-mm grid on the input and output surfaces of the sample. The antennas are the central conductors of copper clad cables, which extend beyond the cladding and are bent by $90^0$ and aligned parallel to the sample's surface. The fields emitted and received at the input and output of the sample are polarized along the direction of the antennas. Because the TM is not complete, perfect transmission is not observed, but ultra-low transmission is measured. New sample configurations are produced by rotating the sample about its axis and then vibrating the sample momentarily so that the sample settles. Field spectra are then reproducible over the 40 hours required for the measurements of the TM of a single realization of the disorder (Supplementary Note 2 and Fig. S3).

**Recursive Green's function simulations**

Simulations of electromagnetic wave propagation are carried out on rectangular samples comprised of a lattice of square boxes with random refractive index. The sides of the boxes have length $a = \frac{\lambda}{2\pi}$ and the refractive index $n = n_r + jn_i$, where the real part $n_r$ is drawn randomly from a rectangular distribution $[1 - \Delta n, 1 + \Delta n]$ with $\Delta n = 0.3$. The imaginary part $n_i$ is calculated from the average decay rate $\gamma$ using the relation $n_i = \gamma n_r \lambda/4\pi$. Each cell has a variance in decay rate due to the variance in $n_i$. Extended Data Fig. 2 represents our computational model. The sample is connected to semi-infinite leads on the left and right and bounded on top and bottom by perfect reflectors. This is equivalent to the sample being uniform and unbounded in the direction perpendicular to the plane of the rectangle. The field excited from the open boundaries of the sample may be expressed in terms of the waveguide modes of the empty waveguide with $n = 1$.

We employ the recursive Green's function method[42] to find the field in the $k^{th}$ column for a source on the left-hand side of the sample. The sample is divided into $K$ columns of width $a$, labeled $1, 2, …, k, … K$. We begin with the surface Green's function of the left lead and proceed from left to right, connecting each column to the subsystem, with columns 1 to $k - 1$ to its left. The coupling between a $k^{th}$ column and the subsystem to the left is expressed via the Dyson equation, $G = G_0 + G_0 V G$. $G$ is the Green's functions of the connected columns from 1 to $k$. $G_0$ is the matrix of Green's functions of the connected columns from 1 to $k - 1$ and the disconnected column $k$, and $V$ is the interaction terms between $k - 1$ and $k$. We ultimately find the result $G_{k1} = [1 - G^L_{kk} V_{k,k+1} G^R_{k+1,k+1} V_{k+1,k}]^{-1} G^L_{k1}$, which allows us to calculate the Green's functions for all columns from 1 to $K$.

The spatial distribution of the output fields, as determined by Green's function, is fitted with a superposition of waveguide modes, and the transmission matrix (TM) is expressed in waveguide mode space. TEs are determined using the singular value decomposition (SVD) of the TM, $t = \mathcal{U} \Lambda \mathcal{V}^\dagger$. Here, $\Lambda$ is a diagonal matrix whose elements are singular values, $\lambda_n = \sqrt{\tau_n}$. $\mathcal{V}$

and $\mathcal{U}$ are unitary matrices on the sample's input and output surfaces, respectively. The amplitude of the $m^{\text{th}}$ channel in the $n^{\text{th}}$ TE on the input and output surfaces are given by the $v_{nm}$ and $u_{nm}$. By projecting the TM onto the input vector $\mathcal{V}$, we obtain the energy density profile at the output surface: $u_n(L) = \sum_{m=1}^{N} |t_{nm} v_{mn}|^2 / v_{wm}$, where $v_{wm}$ is the group velocity of the $m^{\text{th}}$ waveguide mode.

**Impact of spatial resolution on onset of new channels in simulations**

Simulations of the wave equation for the electric field are carried out for a bounded mesh of cells. As a result, the crossover to a new channel does not occur at $N = W/(\lambda/2)$. The difference in the wavelength at the onset of a new propagating channel from $\lambda = W/(N/2)$ decreases as the size of the cells decreases and vanishes in the continuum limit. The calculation of the number of channels and the cause for the difference are outlined below.

For a discretized wave, $k_x^2 = 2\left(1 - \cos(\frac{m\pi}{W})\right) = 4\sin^2(\frac{m\pi}{W})$ and $\sin^2(\frac{k_z}{2}) = \frac{k^2 - k_x^2}{4} = \frac{k^2}{4} - \sin^2\left(\frac{m\pi}{2W}\right)$, where $z$ is the direction of propagation and $x$ is the transverse direction. A sample of $m$ cells in the transverse direction generates a vector of values $\frac{m\pi}{W}$ corresponding to the transverse $k$-vector of propagating modes that enter the system at a particular $\lambda$. But since $k_x < k$ propagating modes must satisfy the condition $\sin^2(\frac{k_z}{2}) = \frac{1}{4}(k^2 - k_x^2) > 0$. Since higher resolution in the simulation corresponds to a larger number of cells of smaller size, simulations at low resolution produce a low-resolution vector of values $\frac{m\pi}{W}$ that generally do not correspond to an integer value of $N$. Extended Data Fig. 3 shows the effects of resolution of the simulation on the accuracy of the value of $N$. As the resolution increases, the value of $N$ rapidly approaches an integer value.

**Extended Data Figures**

**Extended Data Fig. 1 | Fluctuations in conductance in measurements and simulations.**

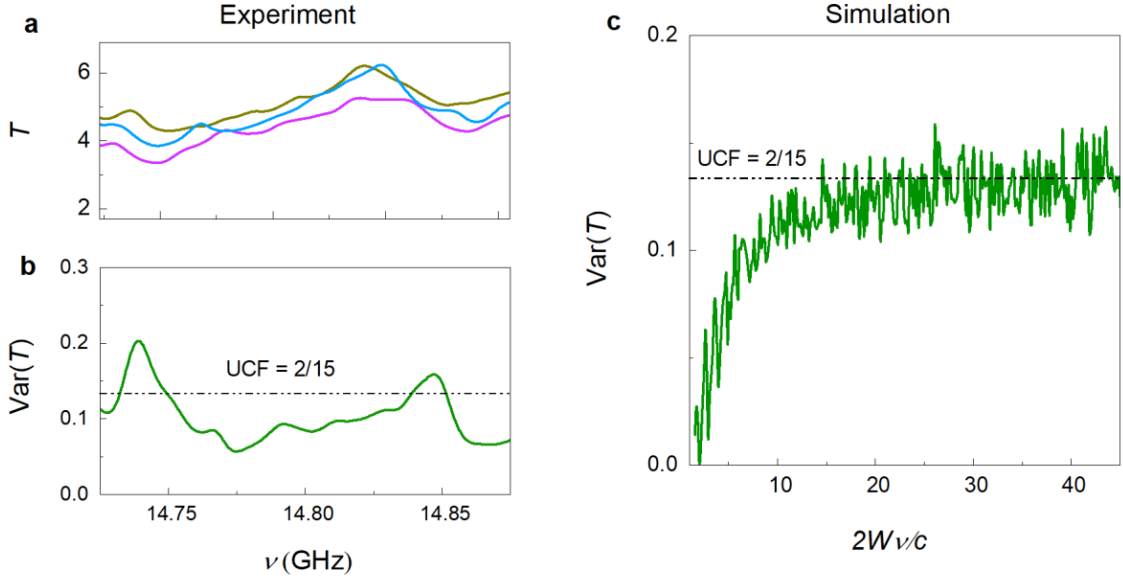

**a,** Measurements of the transmittance in three disordered configurations with $L = 23$ cm. **b,** Measurements of $\text{var}(T)$ for the random ensemble of 23 configurations. The value of $\text{var}(T)$ fluctuates around the value predicted for UCF of 2/15. **c,** Simulations of $\text{var}(T)$ as a function of width at fixed length $L = 1000a$ for the same samples for which results are shown in Fig. 4d. As $W$ increases bringing the sample into the diffusive regime with $g > 1$, $\text{var}(T)$ approaches the predicted value of UCF of 2/15.

**Extended Data Fig. 2 | Computational geometry.**

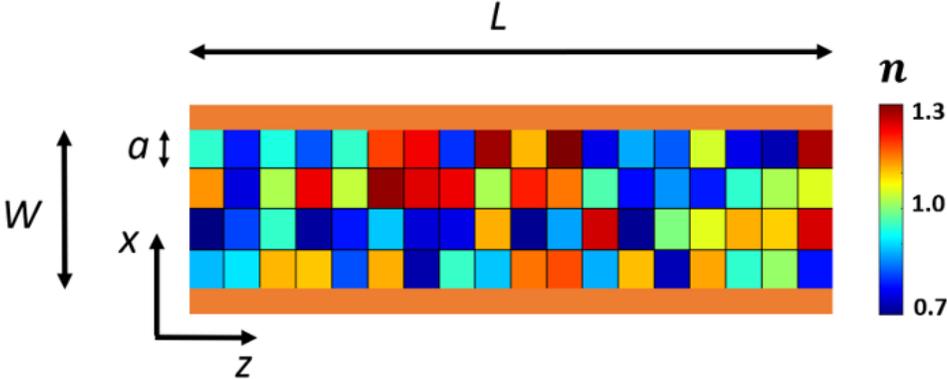

Model sample used in numerical simulations with length $L$, width $W$, and cell size $a$. The sample is bounded with perfect reflectors (orange) along the transverse boundaries in the $z$-direction, is connected to semi-infinite leads at the input and output surfaces along the $x$-direction and is constant in the $y$-direction.

**Extended Data Fig. 3 | Effect of spatial resolution on crossovers.**

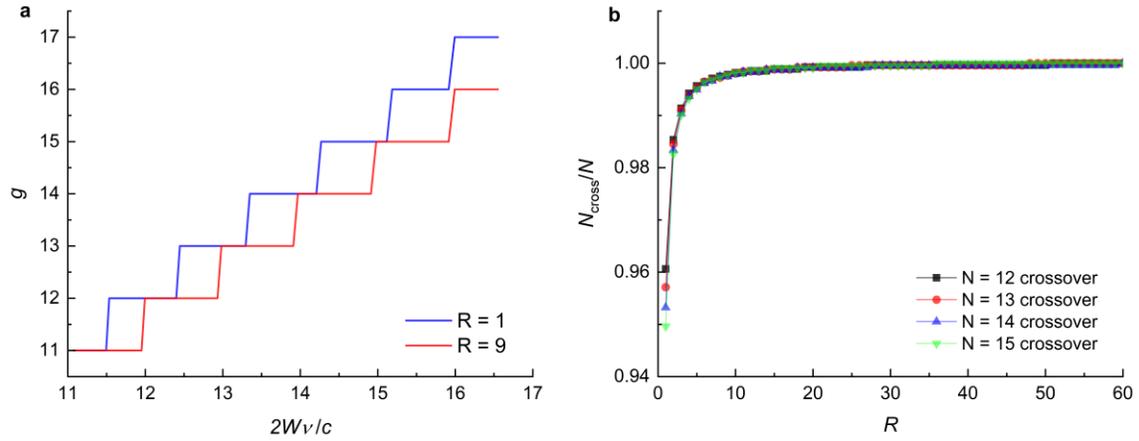

**a,** $N_{\text{cross}}$ is the value of $2Wv/c$ at a crossover as the width is increase in steps of $a$ in an empty waveguide of lengths $100a$ and width $52a$ at two different spatial resolutions, $R = \frac{\lambda}{2\pi a}$. The crossovers approach integer values of $N$ as $R$ increases. **b,** Ratios of $N_{\text{cross}}$ to integer values of $N$ for a crossover from $N$ to $N+1$ versus R. Simulations in this paper were carried out for $R = 1$.

**Extended Data Fig. 4 | Modulation of conductance and the density of states.**

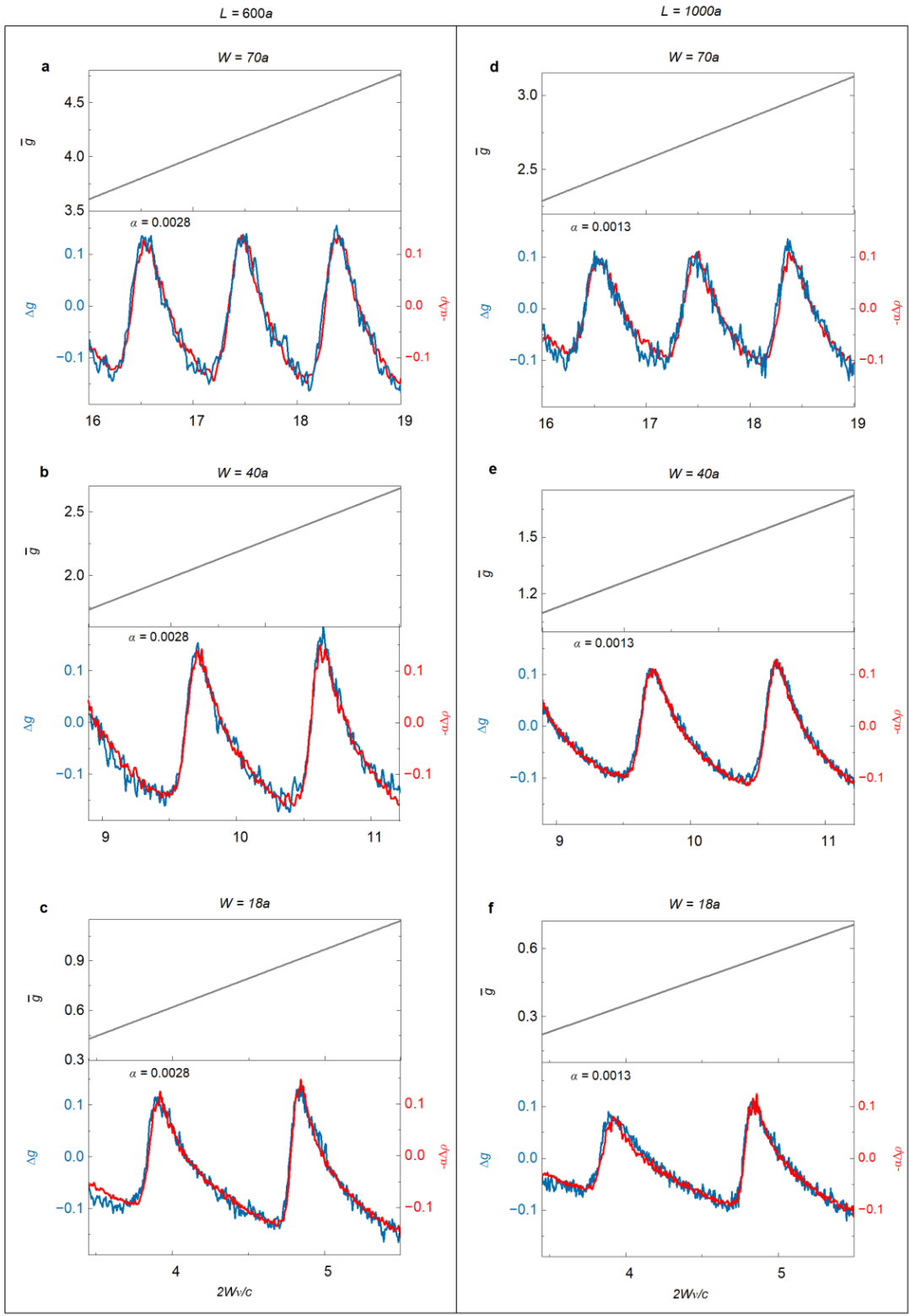

**a-f,** Fits of $-\alpha\Delta\rho$ to $g$ near the channel crossovers. The upper panel gives the slowly varying contribution to the conductance, which is associated with the Thouless conductance. For each length, the proportionality constant, $\alpha$, is independent of sample width. $\alpha$ falls with increasing sample length.

**Extended Data Fig. 5 | Measured conductance and density of states.**

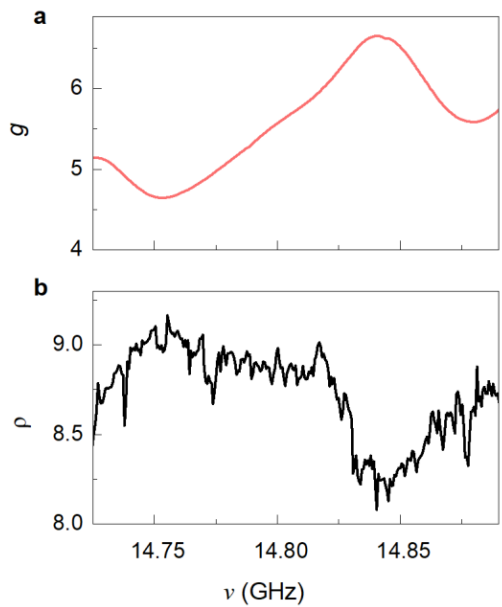

**a,** The measured conductance for $L = 23$ cm, as given in Fig. 1c, and **b,** the corresponding DOS, are compared. There is greater noise in the determination of the DOS, which is obtained from the transmission time, $\rho = \tau_\mathrm{T}/\pi$, than of $g$, because the difference between the phase at two frequencies needs to be determined.

**Extended Data Fig. 6 | Correlation between transmission eigenvalues**.

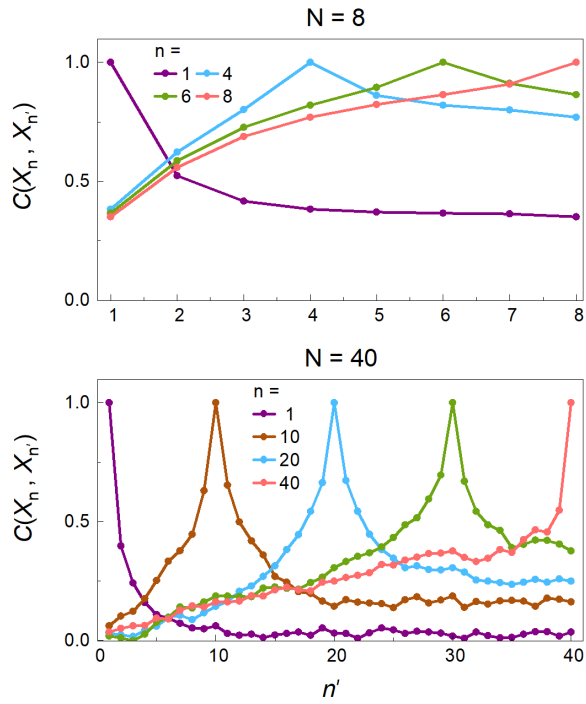

Correlations between various transmission eigenchannels for $N = 8$ (a) and $N = 40$ (b). The correlation between $\tau_N$ and the open channels is still appreciable for $N = 8$, but is significantly attenuated for $N = 40$.

**Extended Data Fig. 7 | Impact of TZs on conductance in single configurations.**

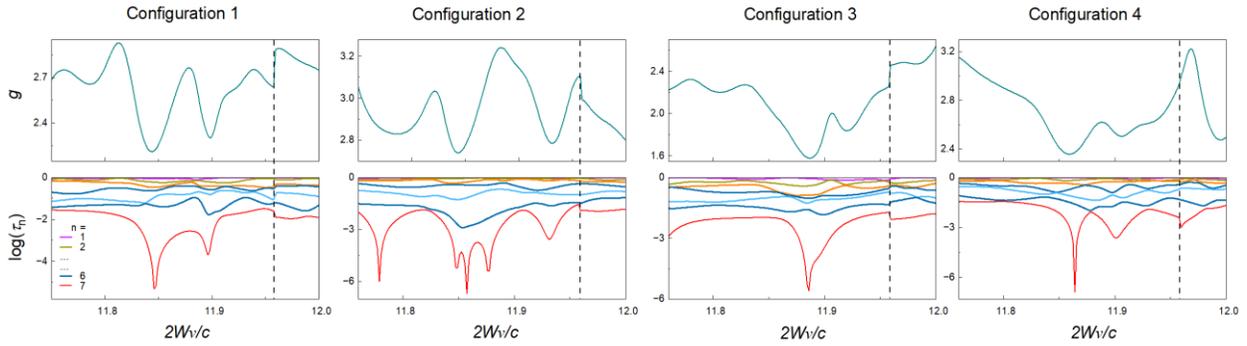

Simulations of spectra of the conductance (top panel) and transmission eigenvalues (bottom panel) for four different configurations with $L = 200a, W = 30a$ and disorder $\Delta n = 0.3$. For isolated TZs, the conductance dips close to the frequency of the TZ.

**Acknowledgements**

We thank Yiming Huang and Asher Maor for valuable discussions. This work is supported by the National Science Foundation (US) under NSF-BSF Award No. 2211646 (AZG).

**Author information**

**Authors and Affiliations**

**Queens College and The Graduate Center of the City University of New York, USA**

K. Joshi, I. Kurtz, Z. Shi and A. Z. Genack

**OFS Labs, USA**

Z. Shi

**Author contributions**

A.Z.G conceived and directed the project, Z.S. measured the microwave transmission matrix. K.J. and I.K. participated in the planning of the research, analysed the experimental data and carried out and analysed numerical simulations. A.Z.G. and J.K., I.K., and Z.S. wrote the manuscript.

**Materials and Correspondence**

Correspondence to A. Z. Genack or K. Joshi, at agenack@qc.cuny.edu or krishnaaj078@gmail.com.

**Ethics Declarations**

The authors declare no competing interests.

# Supplementary Information for "Ohm's law lost and regained: observation and impact of transmission zeros"

Krishna Joshi[1,2], Israel Kurtz[1,2], Zhou Shi[1,2,3] and Azriel Z. Genack[1,2]

[1]Department of Physics, Queens College of the City University of New York, Flushing, New York 11367, USA
[2]Physics Program, The Graduate Center of the City University of New York, New York New York, 10016, USA
[3]OFS Labs, 19 School House Road, Somerset, New Jersey 08873, USA

# INDEX

## Supplementary Notes

**1. Transverse electric and magnetic modes of cylindrical waveguides**

**2. Determination of the experimental signal to noise ratio**

**3. Undoing the impact of absorption in experimental samples**

**4. Transmission time due to transmission zeros with gain and loss**

## Supplementary Figures

**Fig. S1. Measurements of transmission zeros for different lengths**

**Fig. S2. Impact on the conductance of the increasing spectral range of transmission zeros with sample length**

**Fig. S3. Measurement of signal to noise ratio**

**Fig. S4. Undoing the impact of absorption in experimental samples**

**Fig. S5. Lorentzian fit to the transmission zeros**

## References

# Supplementary Notes

## 1. Transverse electric and magnetic modes of cylindrical waveguides

A cylindrical waveguide with radius $a$ and perfect metallic boundary (PMB) conditions supports two modes at each group velocity. The electric field is transverse to the propagation direction for transverse electric modes, while the magnetic field is transverse to the propagation direction for transverse magnetic modes. Due to the rotational symmetry of the waveguide, the solution of Maxwell equations in cylindrical coordinates $(\rho, \varphi)$ can be written as $\psi(\rho, \varphi) = \psi(\rho)e^{in\varphi}, n = 0, 1, 2, \dots$. The solutions of the wave equation for the cylinder is the solution of Bessel's differential equation[1]

**Transverse electric modes:**
$$E_\rho = \frac{j\omega\mu n}{k_c^2 \rho}[A\cos(n\varphi) - B\sin(n\varphi)] J_n(k_c\rho), \quad E_\varphi = \frac{j\omega\mu}{k_c}[A\sin(n\varphi) + B\cos(n\varphi)] J_n'(k_c\rho)$$

**Transverse magnetic modes:**
$$E_\rho = \frac{-j\beta}{k_c}[A\sin(n\varphi) + B\cos(n\varphi)] J_n'(k_c\rho), \quad E_\varphi = \frac{-j\beta n}{k_c^2 \rho}[A\cos(n\varphi) - B\sin(n\varphi)] J_n(k_c\rho)$$

Here, $J_n$ and $J_n'$ are, respectively, the $n^{\text{th}}$-order Bessel function of the first kind and its derivative. The integer $n$ represents the periodicity of the solution in $\varphi$. The PMB boundary conditions at $\rho = a$ give $J_n'(k_c a) = 0$ (transverse electric), $J_n(k_c a) = 0$ (transverse magnetic). If the roots of $J_n$ are defined so that $J_n(p_{nm}) = 0$, where $p_{nm}$ is the root of $J_n$, then $k_c$ must have the value $k_c = p_{nm}/a$. The propagation constant for a transverse magnetic mode with indices $(n, m)$ is given by $\beta_{nm} = \sqrt{k^2 - k_c^2}$, where $k = 2\pi/\lambda$ is the wave number. The condition $\beta_{nm} > 0$ determines the maximum number of propagating modes in the waveguide at a given frequency.

The total number of propagating waveguide modes in our experiments is 62 before the crossover and 64 after the crossover at 14.8317 GHz. Fig. S1 shows spectra of transmission eigenvalues for $n$ = 61, 62, 63, and 64 for lengths $L$ = 23, 40, and 61 cm. The left and right panels show two different sample realizations obtained by rotating the copper tube about its axis. Fig. S2 shows the simulated results of the impact on conductance with sample lengths.

In the experiment, we found a considerable drop in conductance for the crossover N = 61, 62 to 63, 64. However, in our simulated model[2,3], the dip is almost non-existent for N ≥ 40. This behaviour can be attributed to the geometry of our experimental system (3D) and simulated model (2D). The cross-section of a 3D system is larger; hence, the number of propagating modes is greater than that of a 2D system at a given frequency.

## 2. Determination of the experimental signal to noise ratio

Once the measurement of the TM is completed, a measurement of the intensity at the first position and polarization is repeated so that the signal to noise ratio in the experiment can be determined. In Fig. S3, measurements of the first intensity spectrum (blue curve) are compared to spectra taken 40 hours later once the measurement is completed (red curve). The difference in spectra normalized by the mean is shown above each spectrum. The signal to noise ratio is given by $\bar{I}/|\Delta I|$, where $\bar{I}$ is the average of the intensity measured at the start and after the completion of the measurement of the TM, and $|\Delta I|$ is the magnitude of the difference between these signals. The average value of signal to noise ratio is 1300.

## 3. Undoing the impact of absorption in experimental samples

To undo the effects of absorption in our experimental samples (Fig. S4), we apply the following procedure[4–7]:
1. Each data point of the raw spectrum (a, blue curve) is multiplied by a Gaussian cantered at that point (a, black dashed curve).
2. The spectrum is then Fourier transformed into the time domain (b, blue curve).
3. The signal in the time domain is then multiplied by a positive exponential $e^{\gamma t}$, where $\gamma$ is the added gain. The time-domain signal is then cut off at a time delay at which the noise is comparable to the signal (b, red curve) so that noise is not amplified.
4. The signal is then Fourier transformed back to the frequency domain to produce the modified spectrum with added gain (a, red curve).

The probability distribution of intensity obtained in this way are in accord with predictions for a sample without absorption. We note that this process produces a spectrum that is the same as would be obtained from a superposition of modal partial fractions with the linewidth of all modes reduced by $\gamma$. However, because of the nonuniform distribution of gain within the medium, different quasi-normal modes have different rates of decay due to absorption so that absorption cannot be perfectly cancelled by this method.

## 4. Transmission time due to transmission zeros with gain and loss

In the presence of loss or gain, TZs on the real axis in the complex plane are moved off the real axis. If the rate of the loss (gain) of the field of $\gamma$ ($-\gamma$) were the same for all quasi-normal modes, the poles would be displaced vertically in the complex plane by $\gamma$ ($-\gamma$). Since the field is due to the sum over modal contributions, the TZs should be displaced to the same degree. This would give rise to Lorentzian lines in $t_N$ with peaks (dips) of $1/\gamma$ ($-1/\gamma$) and half widths $|\gamma|$, in accord with equation (2). Transmission time $\tau_T$ is the sum of the transmission time of all the TEs, $\tau_T = \sum_n^N t_n^4$. $\tau_T$ is the sum of Lorentzian functions associated, respectively, with the poles, $\tau_p$, and the zeros, $\tau_z$. Hence, each TZ's dips in the $t_n$ will have a Lorentzian line shape. Fig. S5 represents a Lorentzian fit to the 3rd TZ in the simulated spectrum of $t_7$ with the added absorption $\gamma_a = 3 \times 10^{-4}$. The Lorentzian fit function is defined as $L(\omega) = -\zeta_i/[(\omega - Z_i)^2 + \zeta_i^2]$. $Z_i$ and $\zeta_i$ represent the position and half width of the Lorentzian, respectively.

It is evident from Fig. 3e, however, that the dips in $t_N$ are not all equal. This indicates that the narrowing or broadening of quasi-normal modes, and the corresponding displacement of the poles up or down in the complex plane by gain or loss, differ for different modes. The transmission time $t_{N=62}$ for the experimental configuration shown in Figs. 2a-d for the raw data and for three different values of added gain, $\gamma_g$, is shown in Fig. 3f. Gain is added to remove the impact of absorption. The field transmission spectrum is Fourier transformed into the time domain. The time signal is then multiplied by $e^{\gamma t_7}$. The signal at late times is set to zero because the noise is amplified by the added gain (Fig. S4). Dips and peaks occur in $t_{62}$ at the same frequencies at which dips occur in $\log \tau_N$. The variation of spectra of $t_{62}$ with gain or loss shows that these dips are due to TZs near the real axis of the complex plane. The noise in the measurement is greater than in measurements of $\log \tau_N$ because of the greater accuracy needed to calculate the spectral derivative of the field involving difference in the phase of the wave at two frequencies. The accuracy of the determination of the phase derivative from measurements is also compromised by the size of the frequency steps of 300 kHz.

Spectra of the time delay for $N = 62$ are displayed in Fig. 3f for the raw data of the moderately absorbing sample and for added gain of $\gamma_g$ = 0.002 (red), 0.005 (blue), and 0.02 $ns^{-1}$(pink). The four numbered features in these spectra correspond to the four numbered dips in $\log \tau_n$ in Fig. 2b. Features 1, 3, and 4 are negative for the raw data and positive at the highest level of gain. These correspond to single TZs moving from the lower to the upper half of the complex plane as the gain level increases, in accord with the expression for $\tau_z$ in equation (2). In the third numbered feature, the depth of the dip first increases as gain increases before this feature becomes a peak. This indicates that the TZ first moves closer to the real axis as gain is added, and then is pushed above the real axis at the highest level of gain. In the second feature, the raw data shows a peak in the time delay. This corresponds to the case of a conjugate pair of TZs being lowered by absorption, bringing the upper TZ closer to the real axis than the lower TZ[6]. This gives a peak in $t_{62}$ for the raw data which is transformed into a dip as the lower TZ approaches the real axis.

**Fig. S1 | Measurements of transmission zeros for different lengths.**

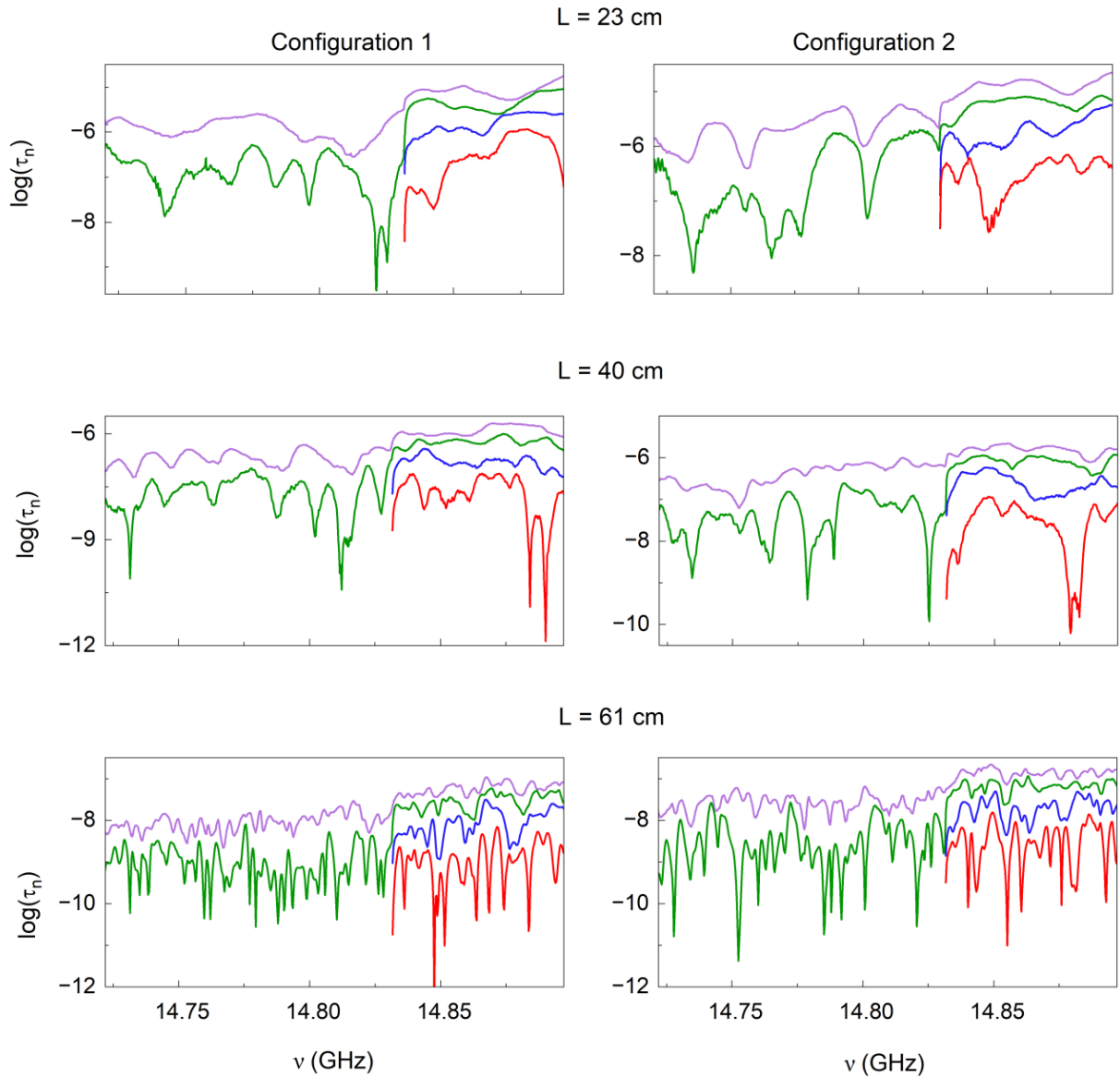

Spectra of the logarithm of transmission eigenvalues for $n = 61, 62, 63,$ and $64$ for sample lengths $L = 23, 40,$ and $61$ cm. The left and right panels present results for two different disordered configurations. The lowest value of $\tau_N$ in the spectra shown is of order of nine. This is nine orders of magnitude below the noise level in the experiment. Aside from noise in the measurement, the lowest value of $\tau_N$ is limited by the closeness of the frequency at which the measurement is made to the frequency of the TZ, and by the nonuniformity of absorption in the medium once the average absorption level is compensated for by adding gain using the approach illustrated in Fig. S4.

**Fig. S2 | Impact on the conductance of the increasing spectral range of transmission zeros with sample length.**

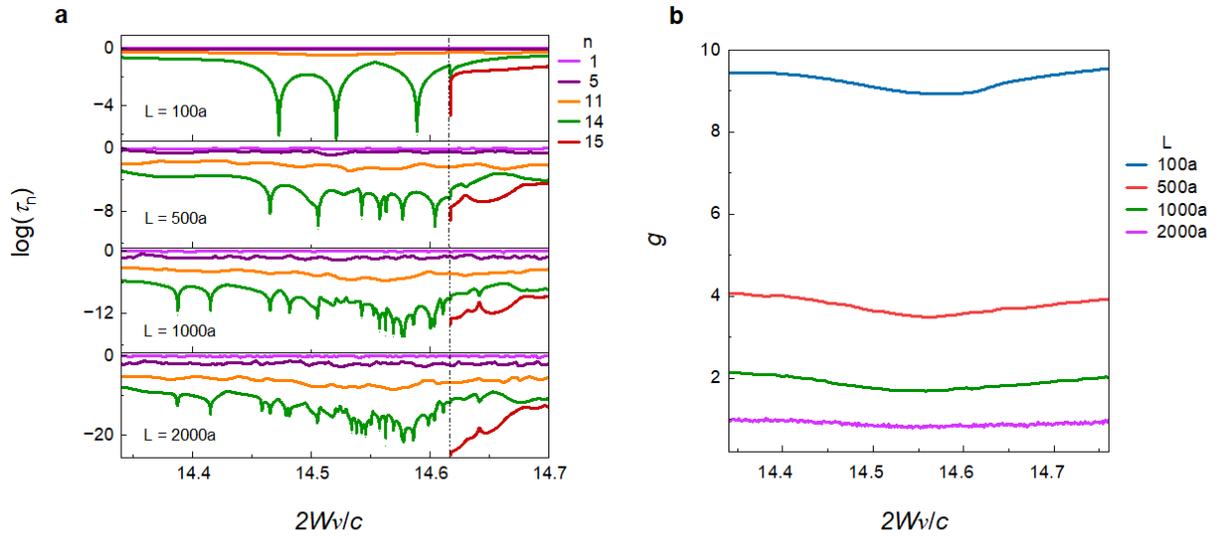

**a,** The spectral range of TZs near the crossover from $N = 14$ to $N = 15$ increases with sample length for individual configurations of width, $W = 60a$. **b,** The dip in the conductance washes out as the length increases as the spectral range of the TZs increases.

**Fig. S3 | Measurement of signal to noise ratio**.

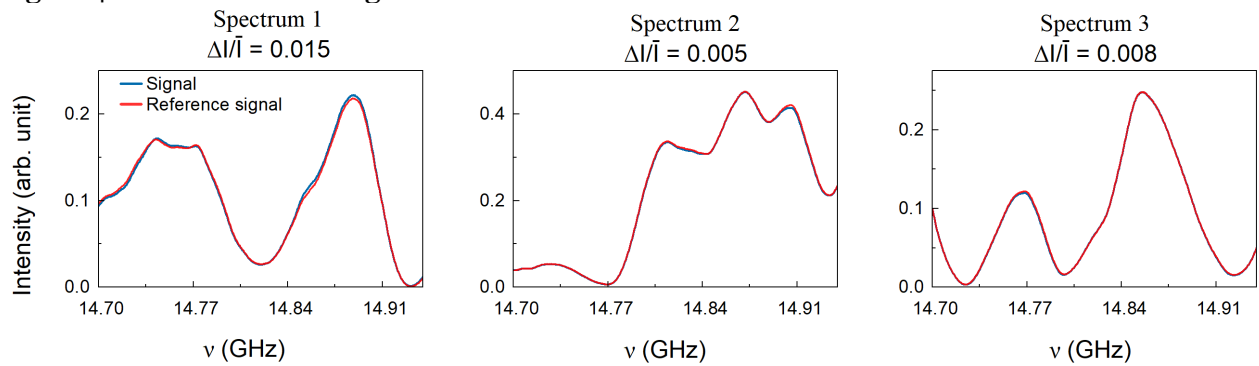

A comparison of the first measured intensity spectrum (blue curve) to the spectrum at the end of the measurement of the TM (red curve) for three different spectra. The difference in spectra normalized by the mean is displayed above each spectrum.

**Fig. S4 | Undoing the effects of absorption in experimental samples.**

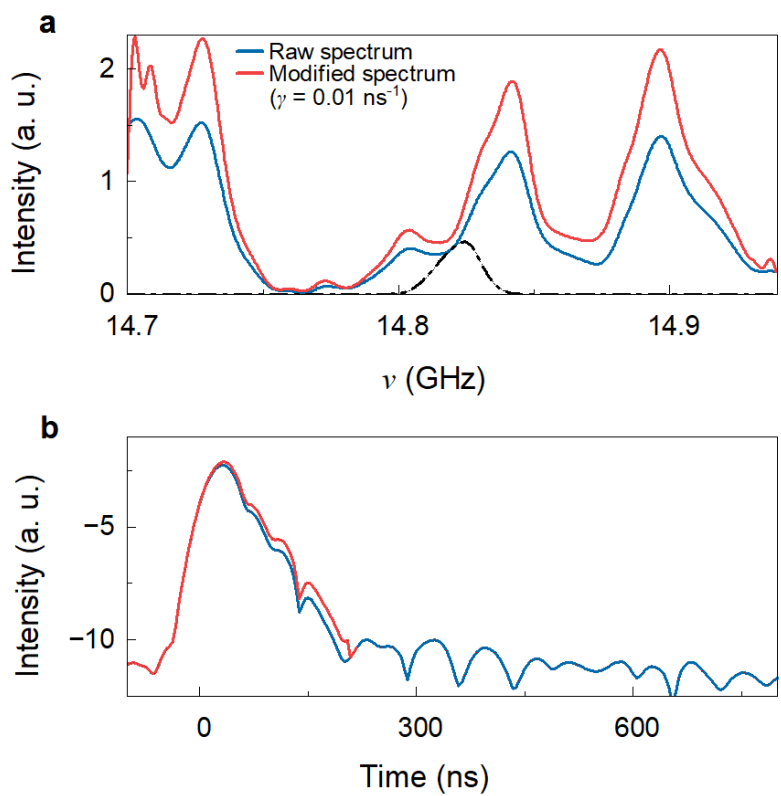

Undoing absorption in the experimental samples: **a,** Raw spectrum (blue curve) and modified spectrum after adding the gain $\gamma_g = 0.01\ ns^{-1}$ (red curve). The dotted black line represents the Gaussian pulse, whose Fourier transform corresponds to the incident pulse whose time response is computed. This makes it possible to act on a small portion of the raw spectrum. **b,** Fourier transform of the raw spectrum (blue curve) and modified spectrum (red curve).

**Fig. S5 | Lorentzian fit to the transmission zeros.**

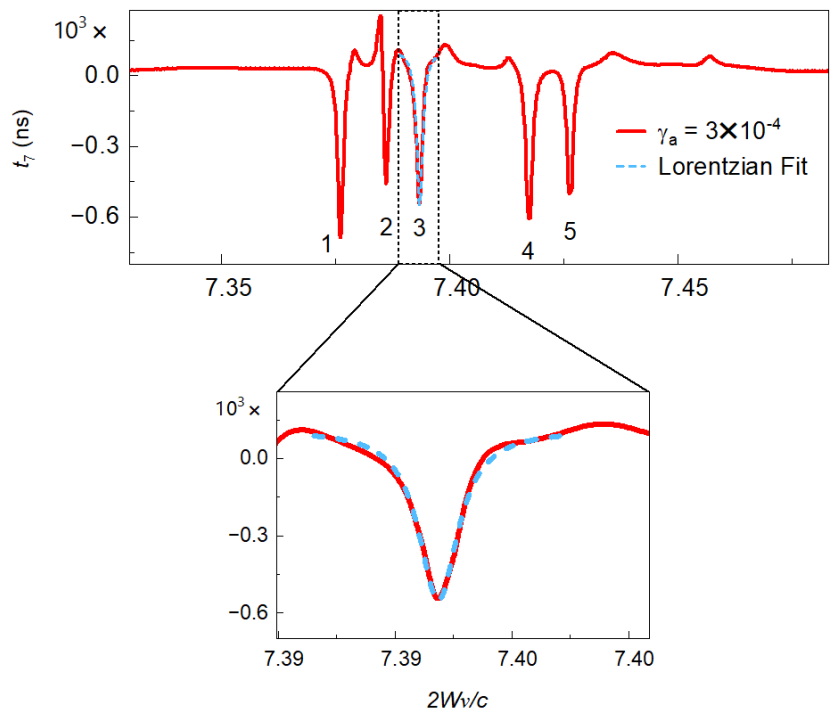

A Lorentzian fit to the 3$^{rd}$ TZ (dotted cyan curve) in the simulated spectrum of $t_7$ (solid red line) with absorption rate $\gamma_a = 3 \times 10^{-4}$.